\documentclass[sigplan,nonacm]{acmart}
\setkeys{acmart.cls}{balance=false}

\setcopyright{none}
\AtBeginDocument{%
}

\usepackage[ruled,vlined,linesnumbered]{algorithm2e}

\usepackage{xspace}
\usepackage{balance}
\usepackage{float}
\usepackage{booktabs}
\usepackage{multirow}
\usepackage{array}
\usepackage{tabularx}
\usepackage{subcaption}
\usepackage{graphicx}
\usepackage{listings}
\usepackage[most]{tcolorbox}
\usepackage{tikz}
\usetikzlibrary{arrows.meta,positioning,calc}

\newcommand{\tool}{\textsc{TyPatch}\xspace}
\newcommand{\knighter}{\textsc{KNighter}\xspace}
\newcommand{\code}[1]{\nolinkurl{#1}}
\newcommand{\figroman}{\rmfamily}

\definecolor{diffgreen}{RGB}{225,247,224}
\definecolor{diffred}{RGB}{255,226,226}
\definecolor{panelblue}{RGB}{226,239,252}
\definecolor{panelorange}{RGB}{255,239,214}
\definecolor{panelgreen}{RGB}{226,245,229}
\definecolor{panelpurple}{RGB}{239,231,250}
\definecolor{panelred}{RGB}{255,232,232}
\definecolor{calloutred}{RGB}{179,38,30}
\definecolor{darkblue}{RGB}{49,95,145}
\definecolor{darkgreen}{RGB}{67,122,70}

\SetAlgoNlRelativeSize{0}
\SetKwInput{KwIn}{Input}
\SetKwInput{KwOut}{Output}
\SetKw{KwRet}{return}
\SetKwFunction{BuildContext}{BuildContext}
\SetKwFunction{GenerateRule}{GenerateRule}
\SetKwFunction{ValidateSchema}{ValidateSchema}
\SetKwFunction{GroundBindings}{GroundBindings}
\SetKwFunction{ValidateStateMachine}{ValidateStateMachine}
\SetKwFunction{ValidateBackend}{ValidateBackend}
\SetKwFunction{NormalizeRule}{NormalizeRule}
\SetKwFunction{SerializeRule}{SerializeTS}
\SetKwFunction{RepairRule}{RepairRule}
\SetKwFunction{BuildCFG}{BuildAnalysisCFG}
\SetKwFunction{JoinAlias}{JoinAliasGraphs}
\SetKwFunction{JoinState}{JoinTypestates}
\SetKwFunction{AliasTransfer}{AliasTransfer}
\SetKwFunction{MatchActions}{MatchActions}
\SetKwFunction{RefutePath}{RefutePath}
\newcommand{\AlgComment}[1]{\textcolor{calloutred}{\bfseries\# #1}}
\newcommand{\TransferArrow}{%
  \par\noindent\makebox[\linewidth][c]{%
    \begin{tikzpicture}
      \draw[-{Stealth[length=2mm,width=1.5mm]},line width=.8pt,darkblue]
        (0,.10) -- (0,-.28);
    \end{tikzpicture}}\par}
\newcommand{\BugLine}[1]{%
  \begin{tcolorbox}[enhanced,frame hidden,boxrule=0pt,colback=diffred,
    left=2pt,right=2pt,top=1pt,bottom=1pt,before skip=0pt,after skip=1pt]
    \footnotesize #1
  \end{tcolorbox}}

\lstdefinestyle{kernel}{
  language=C,
  basicstyle=\ttfamily\footnotesize,
  commentstyle=\ttfamily\upshape,
  columns=fullflexible,
  keepspaces=true,
  showstringspaces=false,
  frame=none,
  numbers=none,
  breaklines=true,
  aboveskip=1pt,
  belowskip=1pt
}

\newtcolorbox{codebox}[1][]{
  enhanced,
  boxrule=.55pt,
  arc=1.5mm,
  colback=white,
  colframe=black!55,
  left=1.3mm,right=1.3mm,top=.9mm,bottom=.9mm,
  before skip=2pt,after skip=2pt,
  #1
}

\newtcolorbox{rqanswer}[1][]{
  enhanced,
  breakable,
  frame hidden,
  boxrule=0pt,
  arc=1.5mm,
  colback=black!10,
  left=2.4mm,right=2.4mm,top=1.5mm,bottom=1.5mm,
  before skip=5pt,after skip=5pt,
  #1
}

\begin{document}

\title[Transforming Patches into Typestate Rules for Kernel Bug Detection]
{\tool: Transforming Patches into Typestate Rules
for Kernel Bug Detection}
\author{Ruoyu Wang}
\authornote{Work done during an internship at Tsinghua University.}
\affiliation{%
  \department{College of AI}
  \institution{Tsinghua University}
  \city{Beijing}
  \country{China}
}
\affiliation{%
  \institution{The University of Hong Kong}
  \city{Hong Kong}
  \country{China}
}
\email{ruoyu\_wang@connect.hku.hk}

\author{Tuo Li}
\affiliation{%
  \institution{Tsinghua University}
  \city{Beijing}
  \country{China}
}
\email{islituo@163.com}

\author{Jia Li}
\authornote{Corresponding author.}
\affiliation{%
  \department{College of AI}
  \institution{Tsinghua University}
  \city{Beijing}
  \country{China}
}
\email{jia\_li@mail.tsinghua.edu.cn}

\renewcommand{\shortauthors}{Wang, Li, and Li}

\begin{abstract}
Historical Linux kernel patches capture defect knowledge that applies beyond
their original repair sites.  Recent work has shown that large language
models (LLMs) can generate static-analysis checkers from historical patches
and use them to uncover new kernel bugs.  However, complete-checker generation requires the
model both to recover the defect semantics expressed by a patch and to
implement sophisticated program-analysis machinery, including object
tracking, alias analysis, path-state maintenance, and interprocedural
propagation.  Coupling these responsibilities in a single end-to-end
code-generation task can turn a simple defect rule into an unstable and
expensive analyzer-implementation problem.

To address this problem, we present \tool, which decouples patch-specific
defect semantics from analyzer implementation.  An LLM translates each patch
into a typestate rule specifying its tracked object, actions, guards,
transitions, and violations.
A shared backend then executes these rules, binding their actions to program
events, tracking object identity across aliases, propagating typestate along
program paths, and producing reports for all rules.  On Linux v6.16,
\tool finds 559 distinct bugs, 121 of which have been confirmed by kernel developers.
In a matched 38-patch comparison with the state-of-the-art
complete-checker construction workflow, \tool uses 88.3--90.1\% fewer
generation tokens, while its initial report pools achieve
3.42--14.95$\times$ the precision of those produced by that workflow.
\end{abstract}



\maketitle

\section{Introduction}

The Linux kernel imposes many subsystem-specific rules on how objects are
created, checked, used, and released.  Violating these rules can cause crashes,
resource leaks, and memory-safety bugs.  The kernel is large and highly
configurable, and many code paths are exercised only with specific hardware or
under rare error conditions.  Consequently, ordinary test workloads cover only
a fraction of the relevant behavior, making bugs in the remaining paths
difficult to uncover through code review and dynamic testing alone.

Static-analysis tools such as Smatch, Coccinelle, and the Clang Static
Analyzer can continuously search large portions of the kernel for
defects~\cite{smatch,padioleau2008coccinelle,clangsa}.  However, the classes of
bugs they detect largely depend on the rules that experts have already
implemented.  For a new long-tail defect pattern, developers must not only
understand the relevant kernel semantics, but also encode those semantics as
an executable program analysis.  As a result, much of the defect knowledge
already present in historical fixes has not been converted into persistent
detection capability for the rest of the kernel.

Historical fixes provide a distinctive source of knowledge for constructing
new static-analysis rules.  A patch records the developer's explanation of
the root cause, the concrete repair action, and the behavioral difference
between the buggy and fixed versions.  The knowledge expressed by a patch is
often not confined to the modified lines; it can capture a transferable
relationship among an object, the operations performed on it, and the
conditions under which those operations are valid.  Automatically extracting
this relationship would allow a single repair to guide automated detection of
similar bugs across the kernel.

\knighter represents the state of the art in automated patch-to-checker
generation.  It uses an LLM to understand Linux repair patches, generate
detection plans, and implement complete Clang Static Analyzer checkers.  This
design combines automatic bug-pattern learning with scalable repository-wide
analysis: the generated checkers have discovered many previously unknown Linux
kernel bugs~\cite{knighter}.

Despite this success, complete-checker generation remains unreliable and
expensive.  \knighter's own failure analysis identifies checker implementation
as its largest source of failure.  Of 22 patch commits that yielded no valid
checker, 13 (59\%) were attributed to inaccurate implementation, compared with
2 to inaccurate bug-pattern analysis and 7 to inaccurate plans~\cite{knighter}.
In our matched 38-patch study, \knighter produces only 27 valid checkers out of
167 candidates with GPT-5.5 and 23 out of 198 with DeepSeek-v4-pro.  Failed
candidates trigger additional generation, compilation repair, and validation.
The two runs consume 7.28 and 9.88 million generation tokens.  A low yield of
valid artifacts leaves much of the recovered defect knowledge unused, while
implementing a separate checker for every patch limits how much repair history
can be turned into persistent detection capability.

\begingroup
\emergencystretch=1em
This fragility stems from coupling patch understanding to analyzer
implementation.  The model first infers the patch-specific
object-state relation and then implements it as a complete checker.
Implementing it requires framework callbacks and state containers for
alias-aware, path-sensitive interprocedural propagation and reporting.  These
analysis mechanisms are difficult but largely
common across patches~\cite{pata,spata}; an implementation error can therefore
invalidate a checker after successful semantic recovery, and each retry
regenerates similar analyzer machinery.\par
\endgroup

To remove this coupling, \tool separates the two responsibilities.  The LLM
generates only the patch-specific defect semantics as a structured typestate
rule, while a shared backend supplies the program-analysis mechanisms for all
accepted rules~\cite{strom1986typestate}.

\begin{figure}[!t]
  \centering
  \includegraphics[width=.90\columnwidth]{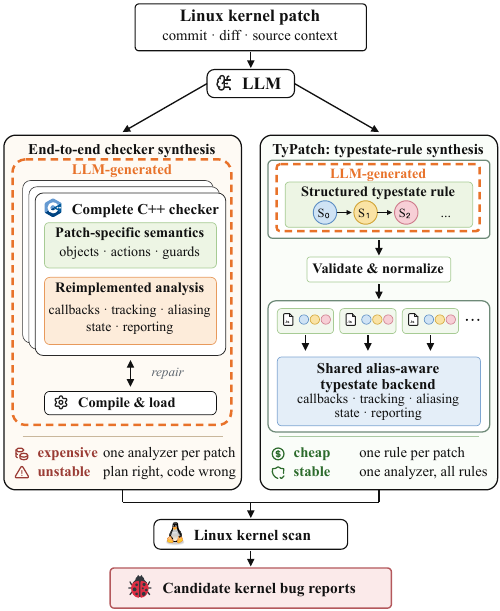}
  \caption{Generate the patch-specific rule; reuse the analyzer.
  Complete-checker generation asks the model to produce both defect semantics
  and checker implementation.  \tool instead generates a structured typestate
  rule and executes it with a shared backend.}
  \Description{Two workflow diagrams compare complete-checker generation with
  TyPatch. In the former, the model must generate both patch-specific defect
  semantics and a checker implementation. In TyPatch, the model generates a
  structured typestate rule, and a shared analyzer provides object tracking,
  path-state propagation, and report construction.}
  \label{fig:generation-boundary}
\end{figure}

Figure~\ref{fig:generation-boundary} summarizes the two-stage pipeline.  First,
for each patch, \tool combines the commit message, code diff, and related source
context as evidence.  An LLM then generates a typestate rule specifying the
tracked object, key actions and guards, state transitions, violation state,
and report evidence.  Deterministic validation and repair then produce a valid
rule or \textsc{NoRule}.  Second,
the shared backend executes the accepted rule pool over the kernel's LLVM IR
and control-flow graph (CFG).  It binds actions to program events, tracks objects and typestate
along paths, and emits bug reports.

Implementing \tool poses two central challenges.  First, the rule
representation must capture different object-state relations and bind them
precisely to program events.  \tool's rule representation explicitly defines
tracked objects, actions and guards, state transitions, and report evidence.  Second,
the backend must execute many heterogeneous rules consistently and
efficiently.  \tool provides shared action matching, alias-aware tracking,
path-sensitive propagation, report construction, and grouped execution.

We select 100 historical Linux fixes across eight bug families and use them as
seeds for rule generation.  The resulting rule pools find 559 distinct bugs in
Linux v6.16, 121 of which have been confirmed by kernel developers.  Given the
same 100 seeds, GPT-5.5, DeepSeek-v4-pro, and Claude Opus 4.8 generate
typestate rules for 91, 74, and 84 patches, respectively.
We compare \tool with \knighter on \knighter's public 38-patch dataset,
using the same model families and Linux v6.16 target.  Across GPT-5.5 and
DeepSeek-v4-pro, \knighter consumes
8.56$\times$ and 10.09$\times$ as many generation tokens, respectively.  On
the resulting initial report pools, \tool achieves 4.67\% and 9.11\% precision,
compared with 0.31\% and 2.66\% for \knighter, under GPT-5.5 and
DeepSeek-v4-pro, respectively.  We further review all 249 generated rules:
236 preserve the core defect relation in their seed fixes, while the remaining mismatches
concentrate in object identity, failure predicates, action binding, and state
topology.

We make the following three contributions:
\begin{itemize}
  \item \textbf{Rule representation.} We introduce a typestate representation
        for patch-derived defect semantics with explicit program bindings,
        separating them from general program-analysis mechanisms.
  \item \textbf{System.} We design and implement \tool, which constructs and
        checks rules from commit messages, code diffs, and source context and
        executes them with a shared alias-aware, path-sensitive backend.
  \item \textbf{Evaluation and impact.} We evaluate three LLMs on 100 Linux
        fixes, compare \tool with \knighter on 38 matched patches, and analyze
        rule fidelity; the resulting rules find 559 distinct bugs in Linux
        v6.16, 121 of which have been confirmed by kernel developers.
\end{itemize}

\section{Motivation}
\label{sec:motivation}

Patch-to-checker generation combines two distinct tasks: recovering the defect
relation expressed by a patch and implementing that relation as a correct
analyzer.  The former requires semantic understanding of the commit evidence;
the latter requires framework-specific callbacks, object identity, alias
handling, and path-state updates.  We use one Linux repair to expose the gap
between them.

\subsection{Correct Plan, Incorrect Checker}

\begin{figure}[!t]
  \centering
  \includegraphics[width=\columnwidth]{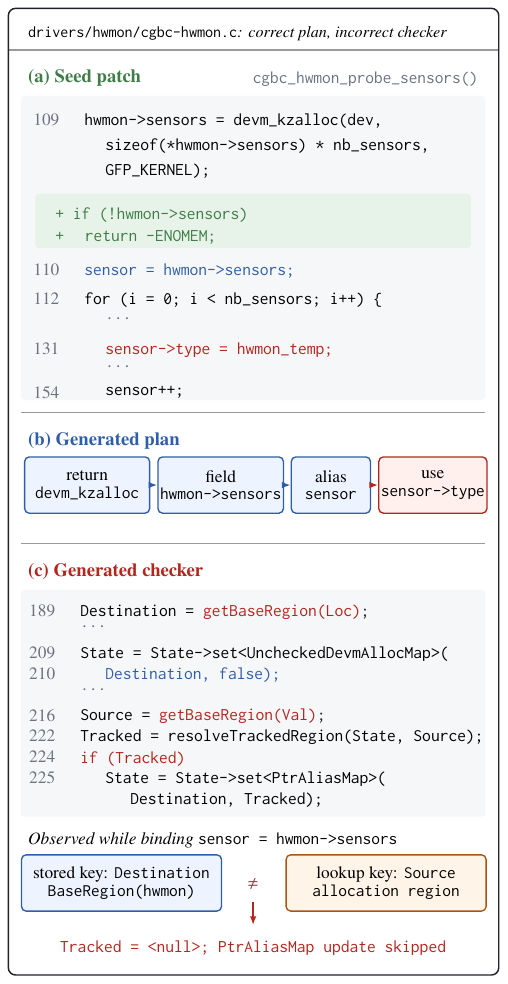}
  \caption{Correct plan, incorrect checker for the \code{cgbc} patch.
  (a) The fix checks the allocation before aliasing and dereference.
  (b) The plan preserves the allocator-to-use chain.  (c) The checker stores
  state under a \code{Loc}-derived key but queries it with a \code{Val}-derived
  key, so the lookup misses and the alias-map update is skipped.}
  \Description{Three-panel example for the cgbc patch. The first panel shows
  an allocation stored in a structure field, copied to a local alias, and
  dereferenced after the upstream fix adds a null check. The second panel
  summarizes the correct generated plan. The third panel shows the generated
  checker using inconsistent region keys, which prevents alias state from being
  propagated.}
  \label{fig:alias-gap}
\end{figure}

Figure~\ref{fig:alias-gap}(a) shows a null-pointer dereference in the
\code{cgbc} hwmon driver and the check later added upstream.  The function
stores the return value of \code{devm_kzalloc()} in
\code{hwmon->sensors}, copies that field into the local cursor
\code{sensor}, and dereferences the cursor while initializing the array.
Because the allocator may return \code{NULL}, the added guard returns before
the local assignment on the failure path.  The continuing path therefore
establishes that the allocation reached through \code{sensor} is non-null.

The \knighter plan in Figure~\ref{fig:alias-gap}(b) recovers this relationship:
the allocator return flows through the structure field and local alias to the
unsafe use.  Implementing that plan as a CSA checker is a different task.  The
checker must preserve one object identity across these representations, attach
the non-null fact only to the continuing branch, and recover the same object
state at the later field access through the appropriate CSA callbacks and
\code{ProgramState} keys.

This distinction appears consistently in the generated artifacts for this
patch.  All 20 generated plans identify the missing null check before the
subsequent dereference, and all 20 generated checkers compile.  Yet only 2 of
the 10 checkers generated with GPT-5.5 and none of the 10 generated with
Opus 4.8 distinguish the buggy seed from its fixed version.  For this patch,
18 of 20 candidates therefore fail
after recovering the missing-check relationship, when that relationship must
be implemented through CSA memory regions, callbacks, and path states.

The generated checker in Figure~\ref{fig:alias-gap}(c) stores the unchecked state using a region derived from
the destination expression, but later queries the alias map using a different
region derived from the bound value.  These expressions resolve to different
CSA regions, so the alias update for \code{sensor} is skipped and the checker
reports neither the buggy nor the fixed seed.  This is not a failure to recover
the patch semantics: the plan already contains the correct allocation, alias
chain, guard, and use.  It arises because complete-checker generation requires
each generated checker to rebuild CSA-specific object and path machinery.  Every
checker must independently select callbacks, region keys, state containers,
and transitions; compilation repair cannot ensure that these choices implement
the recovered plan.  Coupling patch understanding to analyzer implementation
therefore turns a small defect relation into an unstable analyzer-implementation
task.

\subsection{Generate the Rule, Reuse the Analyzer}

This coupling can be avoided by changing the generation boundary.  Across
patches, the tracked object, actions, guards, and transitions vary, whereas
object propagation, alias handling, and path-state maintenance are common
analysis mechanisms.  \tool therefore asks the model to emit only a structured
typestate rule and implements those common mechanisms once in a shared backend.

\begin{figure}[!t]
  \centering
  \includegraphics[width=\columnwidth]{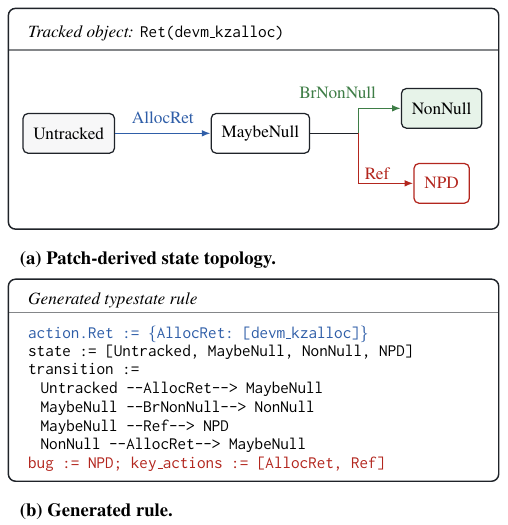}
  \caption{The \code{cgbc} typestate rule and its bindings to
  \code{devm_kzalloc()}, the non-null guard, and the dereference.}
  \Description{State-machine diagram for the cgbc rule. The allocation result
  enters MaybeNull; a successful non-null guard transitions to NonNull; a
  dereference from MaybeNull transitions to the NPD violation state. Labels
  connect the transitions to the allocator return, guard branch, and
  dereference.}
  \label{fig:motivating-rule}
\end{figure}

For the same \code{cgbc} patch, the generated rule binds the source to
\code{devm_kzalloc()}, the guard to its non-null successor, and the violation
to the later dereference.  The backend carries the object through the structure
field and local alias and applies the non-null transition only to the
continuing branch---the mechanics that the generated checker failed to encode.
GPT-5.5 and Opus 4.8 each generated this rule in one attempt;
Figure~\ref{fig:motivating-rule} shows its state topology and action bindings
for this patch.

This example makes this separation concrete.  The next section formalizes
it through \tool's rule representation, construction process, and shared
typestate analysis.

\section{Method}
\label{sec:method}

\noindentparagraph{Overview.}
\tool generates patch-specific typestate rules and executes them with a shared
backend.  Each rule records the tracked object, state-changing actions
and guards, state transitions, and report evidence.  The backend binds rule
actions to program events and tracks objects across aliases.  It propagates
typestate and constructs reports.

\begin{figure}[!t]
  \centering
  \includegraphics[width=\columnwidth]{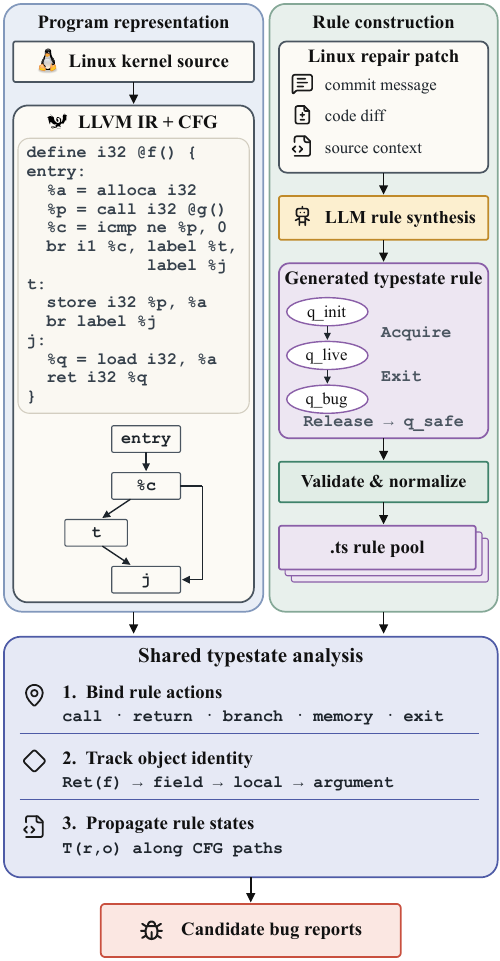}
  \caption{Architecture of \tool.}
  \Description{Pipeline diagram of TyPatch. Commit messages, code diffs, and
  source context feed structured rule generation and validation. Accepted rules
  are loaded by a shared analysis that binds actions to program events, tracks
  aliases, propagates typestate along paths, and emits source-level reports.}
  \label{fig:architecture}
\end{figure}

Figure~\ref{fig:architecture} shows how these components interact.  For each
patch, \tool combines the commit message, code diff, and source context.  It
returns either a validated rule or \textsc{NoRule}.  The backend
loads all accepted rules and analyzes the kernel's LLVM IR and CFG.  The
rule representation connects patch-derived defect knowledge to the shared
backend.  It expresses that knowledge in a form the backend can execute.  We
first define this representation, then describe its construction and execution.

\subsection{Typestate Rule Representation}

The shared backend requires a precise description of the defect relation
recovered from a patch.  The rule representation provides this description as
a state machine with the bindings needed to execute it.  The complete rule is
\[
R=\langle O,E,Q,q_0,\delta,q_b,K\rangle.
\]
Its typestate machine
is \((E,Q,q_0,\delta,q_b)\).  Here, \(E\) is the action alphabet, \(Q\) is a
finite abstract state domain, and \(q_0\in Q\) is the initial state.  The
function \(\delta:Q\times E\rightarrow Q\) defines state transitions, and
\(q_b\in Q\) is the violation state.

The remaining components connect the machine to a program and to a report.
\(O\) specifies how a starting action selects the program value that represents
the tracked object.  The evidence contract is \(K=(\bar e,\Phi)\).
Here, \(\bar e=\langle e_1,\ldots,e_m\rangle\) is the ordered key-action
sequence, and \(\Phi\) contains report constraints such as path consistency or
distinct action locations.  The backend drops a candidate only if it proves a
constraint impossible.  Thus, \(\delta\) describes how matched actions affect
state, whereas \(K\) describes the evidence required for a report.

The domain \(Q\) also has a deterministic join \(J_Q:Q\times Q\rightarrow Q\)
for CFG merges.  For a rule \(r\) with state domain \(Q_r\), we write
\(J_r=J_{Q_r}\) for its join operator.  A generated rule may declare merge cases, and normalization
completes them with a shared fallback policy.  Explicit cases take precedence,
and the violation state is absorbing.  For scalar uninitialized-use rules,
joining an initial and non-initial state yields the initial state; for lifecycle rules,
the same pair preserves the non-initial obligation.  Any remaining unmatched
pair follows the CFG's deterministic predecessor order.  This makes join
behavior an explicit part of the normalized state domain.

\begin{figure}[!t]
  \centering
  \includegraphics[width=\columnwidth]{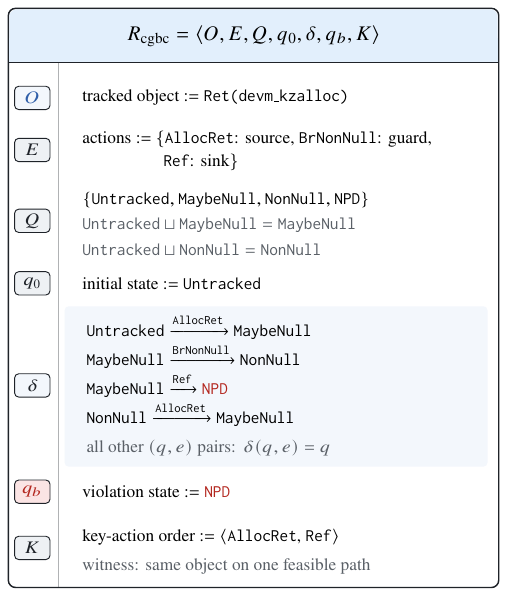}
  \caption{The \code{cgbc} rule instantiated in \(R\).}
  \Description{Semantic decomposition of the cgbc typestate rule into the
  tracked-object binding, action alphabet, state set, initial and violation
  states, transition function, join behavior, and evidence contract. The
  diagram maps serialized rule fields to these components.}
  \label{fig:rule-interface}
\end{figure}

Figure~\ref{fig:rule-interface} gives an instance of \(R\) for the
\code{cgbc} patch in Figure~\ref{fig:alias-gap}(a).  In this rule, \(O\) is the
return value of \code{devm_kzalloc()}, and \(E\) contains the
allocation return, non-null branch, and later dereference.  The transition
function places the allocation in \code{MaybeNull}.  The continuing arm of the
guard changes the state to \code{NonNull}, while an unchecked dereference
reaches \code{NPD}.  A later \textit{AllocRet} resets \code{NonNull} to
\code{MaybeNull} for the new allocation result.  In the figure, \(\sqcup\)
denotes the join operator \(J_Q\).  The two displayed equations are the
rule-specific merge cases; all other state pairs use the shared fallback
policy described above.  The evidence contract \(K\) requires the allocation
and dereference to refer to the same object along a path not proven infeasible.

The bindings attached to \(O\) and \(E\) make the state machine executable.
An object may enter the analysis as a function return, an argument or out
parameter, a field value, or a managed allocation.  A call action names the
callee and selects the argument and optional field that carries the object.  A
return action selects the call result, and a branch action selects an outgoing
CFG edge.  A memory action selects a load, store, or dereference.  An exit
action selects a function return.  For example,
\code{Call[of_node_put,arg0]} denotes a release of the object passed as the
first argument to \code{of_node_put()}; ``release'' alone would not identify a
program event.

The rule omits unbound operations that only transport an object between
state-changing actions.  The shared analysis handles unbound loads, stores,
field address computations, casts, and actual-to-formal transfers.  A memory
operation appears in \(E\) only when the repair assigns it typestate
significance.  A rule specifies its state joins and witness constraints, while
the shared analyzer implements the alias propagation and feasible-path
confirmation needed to enforce them.

\subsection{Constructing Rules from Patches}
\label{sec:construction}

Constructing a rule requires evidence that is rarely stated in one place.  The
commit message states the defect and repair intent.  The diff identifies the
operation that changes behavior.  The surrounding pre-fix source reveals the
object flow and conditions on which the repair depends.  \tool combines these
sources before generation.  When a macro, wrapper, or cleanup declaration
hides the state-changing operation, context construction resolves its
definition.  Each excerpt records its file and line range, which source
grounding later uses to verify APIs named by the generated rule.

The LLM receives this evidence together with the typed rule schema.  It
generates one complete candidate rule.  The candidate names the tracked-object origin,
declares its actions and guards, and supplies the state machine and violation
evidence in a single structured artifact.  It may instead return
\textsc{NoRule} when the repair depends on semantics outside the rule
interface, such as numeric bounds.

The validation pipeline has two stages separated by normalization.  Before
normalization, \emph{schema validation} checks field types, required action
fields, and references to declared states and actions.  \emph{Source
grounding} then checks every function named by a call or return action against
the collected evidence.  Once both checks pass, normalization canonicalizes
the candidate and fills unlisted state--action pairs with identity transitions.

After normalization, \emph{state-machine validation} checks that \(q_b\) is
reachable and that replaying \(\bar e\) from \(q_0\) reaches \(q_b\).
\emph{Backend validation} then requires a source action that identifies the
tracked object, rejects unsupported action combinations, and verifies that the
rule can be translated into the backend representation.  Across the two
stages, argument positions, field paths, and branch or exit kinds are also
checked structurally.  The first stage establishes that a candidate is well
formed and grounded; the second establishes that the normalized rule is
internally consistent and executable.

Validation failures are reported at the rule-field level.  A revision request
contains the original patch evidence and the rejected candidate.  It also
reports diagnostics such as an ungrounded callee, an unreachable violation
state, or a key-action sequence that does not form a transition path.  The
model may revise the candidate twice, and every revision passes through the
complete validation pipeline again.  If no candidate succeeds, construction
returns \textsc{NoRule}.

Once the normalized rule passes state-machine and backend validation, \tool
serializes it in the declarative \code{.ts} format consumed by the shared
analyzer.  Normalization preserves the tracked object, bound actions,
state-changing transitions, violation state, and evidence contract.
Appendix~\ref{app:construction} gives the complete construction algorithm,
context limits, example selection, normalization, and validation details.

\subsection{Shared Typestate Analysis}

A generated rule specifies the program events that change typestate.  Executing
it over LLVM IR requires the backend to (1) bind each action to a concrete
event and selected value, (2) preserve object-specific state across aliases and
control flow while discarding histories proven inconsistent with any path, and
(3) execute heterogeneous rules without mixing their states or evidence.
\tool uses shared action matching and alias-aware tracking to connect rule
actions to program events and objects.  Path-sensitive propagation carries
rule states along the CFG, and evidence screening removes candidates proven
infeasible.  Grouped execution applies this machinery across multiple rules.
Algorithm~\ref{alg:backend} gives the complete workflow.  It joins incoming
facts, transfers object identity, applies matched actions, and screens the
resulting candidates.  We next explain the three mechanisms that support this
workflow.

\begin{algorithm}[!t]
\caption{Shared Typestate Analysis}
\label{alg:backend}
\DontPrintSemicolon
\KwIn{LLVM link unit \(U\), rule set \(\mathcal{R}\)}
\KwOut{Ordered source-level reports \(D\)}
\BlankLine
\AlgComment{Prepare the analysis unit}\;
\(G\leftarrow\operatorname{BuildAnalysisCFG}(U,\mathcal{R})\) by removing loop and return backedges\;
\(W\leftarrow[\,]\); \(D\leftarrow[\,]\)\;
initialize entry-edge states \(\Gamma=\langle A,T,H,B\rangle\)\;
\ForEach{node \(n\) in topological order of \(G\)}{
  \AlgComment{Join predecessor facts}\;
  \(A_n\leftarrow\operatorname{JoinAliasGraphs}(\{A_p:p\in \operatorname{pred}(n)\})\)\;
  \((T_n,H_n,B_n)\leftarrow\operatorname{JoinRuleFacts}
    (\{\Gamma_p:p\in \operatorname{pred}(n)\},A_n,\{J_r\})\)\;
  \AlgComment{Transfer object identity}\;
  \(A_n\leftarrow\operatorname{AliasTransfer}(n,A_n)\)\;
  \(\Gamma_n\leftarrow\langle A_n,T_n,H_n,B_n\rangle\)\;
  \AlgComment{Apply node and exit actions}\;
  \ForEach{\((r,e,v,f)\in\operatorname{MatchActions}(n,\mathcal{R})\)}{
    \(\operatorname{Advance}(\Gamma_n,r,e,v,f,n,W)\)\;
  }
  \AlgComment{Fork successor states and apply edge actions}\;
  \ForEach{outgoing edge \(\ell\) of \(n\)}{
    \(\Gamma_\ell\leftarrow\operatorname{copy}(\Gamma_n)\)\;
    \ForEach{\((r,e,v,f)\in\operatorname{MatchActions}(\ell,\mathcal{R})\)}{
      \(\operatorname{Advance}(\Gamma_\ell,r,e,v,f,\ell,W)\)\;
    }
    store \(\Gamma_\ell\) on \(\ell\)\;
  }
}
\AlgComment{Discard paths proven inconsistent}\;
\ForEach{candidate \(d=(r,o,H)\in W\)}{
  \If{\(\Phi_r=\emptyset\) or \(\neg\RefutePath(G,d,K_r)\)}{
    append \(d\) to \(D\)\;
  }
}
\KwRet \(D\)\;
\end{algorithm}

\paragraph{Binding actions to program events.}
A rule describes state-changing events, whereas the backend operates on LLVM
instructions and CFG edges.  The loader indexes actions by binding form and
resolves each match to an instruction or edge, a selected value, and an
optional field path.  Return actions select call results; call actions select
arguments or out-parameter slots; and memory actions select loads, stores, or
dereferences.  Guard actions bind to outgoing edges, and exit actions are
projected over active tracked objects.  All rules share this matcher while
supplying different bindings and transition tables.

\(\operatorname{MatchActions}(\ell,\mathcal{R})\) returns
\((r,e,v,f)\) for actions bound to LLVM instructions or outgoing CFG edges.

\paragraph{Preserving object identity.}
Matched actions belong to one rule execution only when their selected values
represent the same object.  The AliasGraph
\(A=\langle N,F,\rho\rangle\) makes this identity explicit: \(N\) contains
abstract objects, \(F\) contains labeled pointee and field edges, and
\(\rho:V_{\mathrm{IR}}\rightharpoonup N\) maps LLVM values to objects.
Resolving \((v,f)\) obtains \(\rho(v)\) and follows field path \(f\).

Stores replace a destination's \code{ref} edge, loads follow it, and
\code{getelementptr} creates or follows field edges.  Casts preserve nodes;
\(\phi\) and \code{select} do so only when all incoming values resolve to one
node.  At resolved calls, actuals and formals share nodes, allowing stores
through out parameters to remain visible after return.  Summaries similarly
connect recognized driver-data setter/getter pairs.

At CFG joins, predecessor nodes containing the same LLVM value are unified,
followed recursively by destinations reached through the same field label.
Rule slots are remapped to the unified nodes before their states are joined.
State therefore survives field, local-alias, and interprocedural transfers
without encoding those transfers in each rule.

\paragraph{Preserving path-specific state.}
Object identity alone is insufficient when a guard or discharge applies on
only one branch.  \tool removes loop and return backedges and traverses the
resulting acyclic interprocedural graph in topological order.  Each analysis
edge carries \(\Gamma=\langle A,T,H,B\rangle\): the AliasGraph \(A\),
rule-indexed object states \(T\), supporting histories \(H\), and branch facts
\(B\).  One object can thus carry independent state slots for several rules.

For rule \(r\), \(\delta_r\), \(q_b^r\), \(J_r\), and
\(K_r=(\bar e_r,\Phi_r)\) denote its transition function, violation state,
join, and evidence contract.  On a matched action,
\(\operatorname{Advance}(\Gamma,r,e,v,f,\ell,W)\) resolves the selected object,
appends the occurrence to its history, and applies \(\delta_r\).  A starting
action first initializes the object's rule slot at \(q_0^r\); a projected exit
advances every active object.  Reaching \(q_b^r\) adds a candidate to \(W\).
Exact transitions take precedence over normalized identity self-loops.

Node actions update state before propagation.  The analyzer copies the state
to each successor and applies guard actions only to their bound edges.  At CFG
joins, it merges incoming typestates using \(J_r\).  With conditional-state
tracking enabled, it can retain separate states for complementary null and
non-null branches on the same value until a later guard selects one or
\(J_r\) merges them.  Each retained state carries one representative history.

Since merged facts may originate from different paths, reaching \(q_b^r\)
produces a candidate report.  When \(\Phi_r\) requests screening, the backend
replays the key-action sequence \(\bar e_r\) along individual paths in the
relevant CFG slice.  Each path maintains its own alias graph, typestate, and
branch facts during replay.  This checks whether the required actions can
operate on the same object in order along one path, rather than relying on
the merged scan state alone.  The backend rejects a candidate only if it
proves that no path satisfies the constraints in \(K_r\), including field
paths and distinct action locations.  Unknown cases are retained.  Reports
record the rule, object, bound actions, history, and source sites.

The same traversal executes compatible rules together in one link unit.  They
share one AliasGraph, while every object carries independent rule-indexed
states and histories.  Action indices dispatch each event to all matching
rules after paying its object-transfer cost once.  Adding a rule therefore
does not require another checker implementation or another model call during
the scan.

We now trace the \code{cgbc} example in Figures~\ref{fig:alias-gap}(a)
and~\ref{fig:motivating-rule} through Algorithm~\ref{alg:backend}.  The
allocation action initializes the return from
\code{devm_kzalloc()} in \emph{MaybeNull}.  Alias transfer then carries the
same object through \code{hwmon->sensors} to \code{sensor}.  In the fixed
seed, the non-null edge changes its state to \emph{NonNull}, while the null
edge returns before the dereference.  In the pre-patch seed, no guard performs
this transition.  The later dereference therefore changes \emph{MaybeNull} to
\emph{NPD} and creates a candidate.  Evidence screening retains it because the
allocation, alias transfers, and dereference form a feasible same-object path.

\section{Evaluation}
\label{sec:evaluation}

\noindent We explore the following research questions for \tool:

\noindent\textbf{RQ1.} Can \tool find real-world Linux kernel bugs?

\noindent\textbf{RQ2.} How do rule and checker construction compare in artifact
yield, generation cost, and initial report quality?

\noindent\textbf{RQ3.} How faithfully do rules preserve seed semantics?

\subsection{Experimental Setup}

\paragraph{Datasets and models.}
All scans target Linux v6.16 at commit \code{98a11dadac64}.  RQ1 and RQ3 use
100 historical Linux fixes across eight bug families, with rules generated
independently by GPT-5.5, DeepSeek-v4-pro, and Claude Opus 4.8.  RQ2 compares
both systems with GPT-5.5 and DeepSeek-v4-pro on the 38 commits in
\knighter's public benchmark spanning six shared bug families.

\paragraph{Environment.}
We performed all kernel-wide scans on a dual-socket server with two 12-core Intel Xeon
Silver 4310 CPUs, 256~GB RAM, and Ubuntu 24.04, using eight workers.

\subsection{RQ1: Linux Bug Finding}

Across the three RQ1 rule pools, \tool finds 559 distinct bugs in Linux v6.16,
121 of which have been confirmed by kernel developers.  Before cross-model
deduplication, GPT-5.5, DeepSeek-v4-pro, and Opus 4.8 contribute 319, 384, and
429 TP components, respectively, for a total of 1,132.  The findings span all
eight evaluated typestate families across networking, media, sound, GPU, PHY,
and IIO.

Given the same 100 seed fixes, GPT, DeepSeek, and Opus generate rules for 91,
74, and 84 patches, respectively.  Table~\ref{tab:rq1coverage} breaks this
coverage down by family.  Missing-release, refcount-imbalance, and
acquire--release-order seeds translate consistently across models.  The
largest coverage differences occur for uninitialized-use, use-after-release,
and publish-before-initialization seeds.

\begin{table}[t]
\caption{Generated typestate rules for RQ1's 100 patches.}
\label{tab:rq1coverage}
\centering
\footnotesize
\begin{tabular}{@{}lrrrr@{}}
\toprule
Bug family & Input & GPT & DS & Opus \\
\midrule
Null-Pointer Dereference & 20 & 18 & 17 & 19 \\
Double Action & 20 & 19 & 12 & 15 \\
Missing Release & 17 & 17 & 15 & 17 \\
Refcount Imbalance & 16 & 16 & 15 & 16 \\
Uninitialized Use & 10 & 8 & 8 & 4 \\
Use After Release & 10 & 7 & 3 & 7 \\
Acquire--Release Order & 4 & 4 & 4 & 4 \\
Publish Before Initialization & 3 & 2 & 0 & 2 \\
\midrule
\textbf{Total} & \textbf{100} & \textbf{91} & \textbf{74} & \textbf{84} \\
\bottomrule
\end{tabular}
\end{table}

\begin{figure}[!t]
  \centering
  \includegraphics[width=\columnwidth]{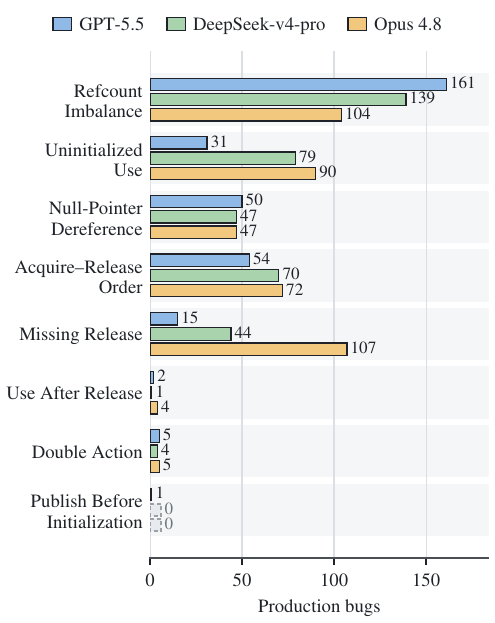}
  \caption{RQ1 true-positive components by bug family; gray stubs denote zeros.}
  \Description{Grouped chart of true-positive components by typestate family
  for the GPT-5.5, DeepSeek-v4-pro, and Opus 4.8 rule pools. Refcount imbalance
  is the largest group for GPT-5.5 and DeepSeek-v4-pro, while missing release is
  more prominent for Opus 4.8. Dashed stubs indicate zero values.}
  \label{fig:rq1families}
\end{figure}

Figure~\ref{fig:rq1families} shows the distribution of these model-specific TP
components.  Refcount imbalances are the largest group for GPT and DeepSeek,
while Opus contributes substantially more missing-release findings.  In our
taxonomy, \emph{Refcount Imbalance} denotes unmatched get/put-style
references; \emph{Missing Release} denotes an unreleased acquired resource.

Of the 559 distinct bugs, 196 are found by all three model-derived rule pools,
181 by exactly two, and 182 by one.  Thus, 377 of 559 bugs (67.4\%) are
surfaced by rule pools from at least two models, while each model also
contributes findings absent from the other two.  Table~\ref{tab:rq1reports}
reports the corresponding model-specific component pools and their
source-adjudicated precision.

\begin{table}[t]
\caption{Model-specific RQ1 report pools after deduplication.  Unique TPs are
not found by the other two models.}
\label{tab:rq1reports}
\centering
\footnotesize
\begin{tabular*}{\columnwidth}{@{\extracolsep{\fill}}lrrrrr@{}}
\toprule
Model & Reports & TP & Unique TP & FP & Precision \\
\midrule
GPT-5.5 & 3,662 & 319 & 54 & 3,343 & 8.71\% \\
DeepSeek-v4-pro & 4,842 & 384 & 40 & 4,458 & 7.93\% \\
Opus 4.8 & 5,655 & 429 & 88 & 5,226 & 7.59\% \\
\bottomrule
\end{tabular*}
\end{table}

Constructing each model-specific rule pool consumes 2.52--3.12 million model
tokens, including repair attempts; Appendix~\ref{app:additional-results}
provides the input/output breakdown.

\begin{figure}[!t]
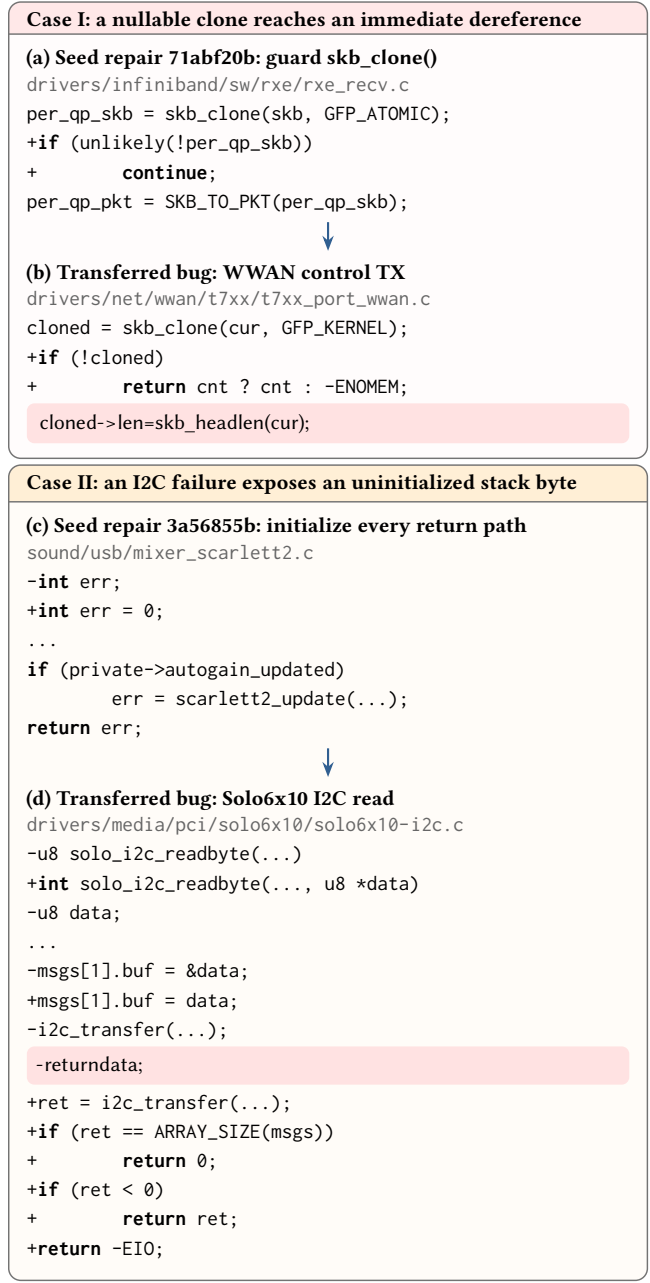

\centering
\begingroup\figroman
\begin{codebox}[title={Case I: a nullable clone reaches an immediate dereference},
  colback=panelred!14,colbacktitle=panelred,coltitle=black,
  fonttitle=\figroman\bfseries\footnotesize]
{\footnotesize\bfseries (a) Seed repair \code{71abf20b}: guard \code{skb_clone()}\par}
{\footnotesize\color{black!60}
\path{drivers/infiniband/sw/rxe/rxe_recv.c}\par}
\begin{lstlisting}[style=kernel]
per_qp_skb = skb_clone(skb, GFP_ATOMIC);
+if (unlikely(!per_qp_skb))
+        continue;
per_qp_pkt = SKB_TO_PKT(per_qp_skb);
\end{lstlisting}
\TransferArrow
{\footnotesize\bfseries (b) Transferred bug: WWAN control TX\par}
{\footnotesize\color{black!60}
\path{drivers/net/wwan/t7xx/t7xx_port_wwan.c}\par}
\begin{lstlisting}[style=kernel]
cloned = skb_clone(cur, GFP_KERNEL);
+if (!cloned)
+        return cnt ? cnt : -ENOMEM;
\end{lstlisting}
\BugLine{\code{cloned->len = skb_headlen(cur);}}
\end{codebox}
\begin{codebox}[title={Case II: an I2C failure exposes an uninitialized stack byte},
  colback=panelorange!16,colbacktitle=panelorange,coltitle=black,
  fonttitle=\figroman\bfseries\footnotesize]
{\footnotesize\bfseries (c) Seed repair \code{3a56855b}: initialize every return path\par}
{\footnotesize\color{black!60}\path{sound/usb/mixer_scarlett2.c}\par}
\begin{lstlisting}[style=kernel]
-int err;
+int err = 0;
...
if (private->autogain_updated)
        err = scarlett2_update(...);
return err;
\end{lstlisting}
\TransferArrow
{\footnotesize\bfseries (d) Transferred bug: Solo6x10 I2C read\par}
{\footnotesize\color{black!60}
\path{drivers/media/pci/solo6x10/solo6x10-i2c.c}\par}
\begin{lstlisting}[style=kernel]
-u8 solo_i2c_readbyte(...)
+int solo_i2c_readbyte(..., u8 *data)
-u8 data;
...
-msgs[1].buf = &data;
+msgs[1].buf = data;
-i2c_transfer(...);
\end{lstlisting}
\BugLine{\code{-return data;}}
\begin{lstlisting}[style=kernel]
+ret = i2c_transfer(...);
+if (ret == ARRAY_SIZE(msgs))
+        return 0;
+if (ret < 0)
+        return ret;
+return -EIO;
\end{lstlisting}
\end{codebox}
\endgroup
\caption{Two seed-to-bug transfers.  Panels (a,b) transfer a nullable-object
rule from RDMA receive to WWAN transmit; panels (c,d) transfer an all-path
initialization rule from ALSA control logic to a media-driver I2C helper.  In
each pair, the right panel shows the detected bug and its corresponding fix;
red bands mark the unsafe sinks.}
\Description{Four code panels arranged as two seed-to-bug transfers. The first
pair shows a nullable clone repair in an RDMA driver and the analogous missing
check in a WWAN driver. The second pair shows an all-path initialization repair
in ALSA and the analogous uninitialized return in a media I2C helper. Arrows
connect each seed repair to the transferred finding, and shaded bands mark
unsafe sinks.}
\label{fig:productioncases}
\end{figure}

\paragraph{Case studies.}
Figure~\ref{fig:productioncases} presents two representative transfers from
historical seed repairs to developer-confirmed findings.  The first carries a
nullable-object check from RDMA to WWAN; the second carries an all-path
initialization requirement from ALSA to a media I2C helper.

\paragraph{Cross-subsystem nullable allocation.}
The seed in Figure~\ref{fig:productioncases}(a) checks the return from
\code{skb_clone()} before an RDMA receive path converts it into packet state.
The resulting nullable-object rule exposes the same unchecked return in the
WWAN control transmit path in panel (b), where allocation failure reaches
\code{cloned->len} under different surrounding control flow.  The accepted fix
adds the missing guard while preserving the driver's partial-write return
convention.

\paragraph{Cross-stack output initialization.}
The ALSA seed in Figure~\ref{fig:productioncases}(c) initializes a returned
status on every path.  The resulting rule finds a media-driver variant in panel
(d): \code{solo_i2c_readbyte()} ignores the number of messages completed by
\code{i2c_transfer()} and returns \code{data} even when the transfer is short.
This can expose an uninitialized stack byte to chip detection, V4L2
input-status queries, and ALSA gain controls.

\begin{rqanswer}
\textbf{Answer to RQ1.}
\tool finds 559 distinct bugs in Linux v6.16, including 121 confirmed by
kernel developers.  These bugs span all eight evaluated typestate families
and multiple kernel subsystems; 377 (67.4\%) are found by rules generated by
at least two models.
\end{rqanswer}

\subsection{RQ2: Rule vs. Checker Construction}
\label{sec:rq2}

We run \tool as described in Section~\ref{sec:construction} and
\knighter with its published default configuration.
\emph{Candidates} count generated rule or checker versions, including repairs;
\emph{Artifact yield} counts input patches whose workflow produces a rule or
checker accepted by its construction pipeline and included in the kernel scan.
Generation-token cost includes all construction attempts.  We execute every
resulting artifact and report the TP, FP, and precision of the
\emph{initial report pools} before optional report-level post-processing.

\begin{table}[!ht]
\caption{Matched-38 artifact construction and initial report quality.  Token
counts are in thousands (K).}
\label{tab:matched38}
\centering
\footnotesize
\begin{subtable}{\columnwidth}
\centering
\caption{Artifact construction cost.}
\label{tab:matched38-cost}
\begin{tabular*}{\columnwidth}{@{\extracolsep{\fill}}llrrrrr@{}}
\toprule
Model & System & Can./Art. & Calls & Input & Output & Total \\
\midrule
\multirow{2}{*}{GPT-5.5}
 & \tool & 57/31 & 88 & 769.6 & 81.2 & 850.8 \\
 & \knighter & 167/27 & 596 & 5,958.0 & 1,325.9 & 7,284.0 \\
\midrule
\multirow{2}{*}{DeepSeek}
 & \tool & 62/25 & 90 & 660.5 & 318.5 & 979.0 \\
 & \knighter & 198/23 & 826 & 5,610.1 & 4,271.6 & 9,881.7 \\
\bottomrule
\end{tabular*}
\end{subtable}

\medskip
\begin{subtable}{\columnwidth}
\centering
\caption{Initial report quality.}
\label{tab:matched38-quality}
\begin{tabular*}{\columnwidth}{@{\extracolsep{\fill}}llrrr@{}}
\toprule
Model & System & Reports & TP & Precision \\
\midrule
\multirow{2}{*}{GPT-5.5}
 & \tool & 4,584 & 214 & 4.67\% \\
 & \knighter & 4,804 & 15 & 0.31\% \\
\midrule
\multirow{2}{*}{DeepSeek}
 & \tool & 1,735 & 158 & 9.11\% \\
 & \knighter & 2,178 & 58 & 2.66\% \\
\bottomrule
\end{tabular*}
\end{subtable}
\end{table}

\paragraph{Artifact construction.}
Table~\ref{tab:matched38}(a) shows the same result under both models: \tool
completes construction for more patches while making fewer model calls and
consuming fewer tokens.  Under GPT-5.5, it generates 57 rule candidates and
constructs rules for 31 of 38 patches with 88 calls and 850.8K tokens;
\knighter generates 167
checker candidates and obtains 27 valid checkers with 596 calls and 7.28M
tokens.  Under DeepSeek-v4-pro, \tool generates 62 candidates and constructs
rules for 25 patches with 90 calls and 979.0K tokens, whereas \knighter
generates 198 candidates and obtains 23 valid checkers with 826 calls and
9.88M tokens.
Thus, \knighter makes 6.8--9.2$\times$ as many model calls, while the
implemented \tool workflow uses 88.3--90.1\% fewer generation tokens.

Across both models, \tool completes construction for more patches than
\knighter at a much lower average generation cost per artifact.  Under
GPT-5.5, the average costs are 27.4K tokens for \tool and 269.8K for \knighter; under
DeepSeek-v4-pro, they are 39.2K and 429.6K, respectively. 

Non-artifacts are concentrated in patches that require reasoning beyond a
sequential object-state relation.  Under GPT-5.5, six of seven non-artifacts
are heterogeneous misuse patches: three require size or bounds reasoning, two
encode a constant or API change without a temporal object relation, and one
fails schema validation.  The remaining UAF candidate omits the use action
needed to form a violation.  DeepSeek shows a similar pattern for value-,
bounds-, and nonprotocol misuse seeds.

\paragraph{Initial report quality.}
Table~\ref{tab:matched38}(b) reports the quality of the initial report pools.
Under GPT-5.5, \tool produces 214 TP instances among 4,584 reports (4.67\%),
compared with 15 among 4,804 reports (0.31\%) for \knighter.  Under
DeepSeek-v4-pro, the corresponding results are 158 of 1,735 reports (9.11\%)
and 58 of 2,178 reports (2.66\%).  The lower precision under GPT-5.5 is
concentrated in three broad rules: a sequential approximation of a concurrent
UAF seed produces 1,753 reports, while two uninitialized-data rules produce
545 and 431.  Together, these rules account for 2,729 (95.8\%) of the 2,849
additional reports produced with GPT-5.5 relative to DeepSeek-v4-pro.  Across
the two models, \tool therefore produces 2.72--14.27$\times$ as many TP instances at
3.42--14.95$\times$ the precision.

\begin{rqanswer}
\textbf{Answer to RQ2.}
Across the matched workflows, \tool produces more artifacts under both models
and uses 88.3--90.1\% fewer generation tokens.
Before report-level post-processing, it produces 2.72--14.27$\times$ as many
true-positive report instances and achieves 3.42--14.95$\times$ the precision
of the corresponding \knighter results.
\end{rqanswer}

\subsection{RQ3: Rule Fidelity and Misalignment}

We define a rule as aligned with its seed patch when it preserves the tracked
object and the source, discharge, and sink roles that constitute the defect
relation.  Generalizations of predicates or API bindings are also considered
aligned if they preserve this core relation.  We apply this definition to all
249 generated rules by comparing each rule with its seed commit message, code
diff, and relevant source context.

Table~\ref{tab:alignment}(a) shows that 236 of 249 generated rules preserve
their seed relation: 88 of 91 for GPT-5.5, 70 of 74 for DeepSeek-v4-pro, and 78
of 84 for Opus 4.8.  All three models preserve the seed relation in more than
nine out of ten generated rules, indicating that current LLMs can reliably
translate patch evidence into \tool's structured rule representation.

Among the 236 aligned rules, we further distinguish exact reproduction from
useful generalization.  Of these rules, 159 preserve every audited object,
action, predicate, scope, and topology field, while 77 broaden at least one
predicate or API binding without changing the tracked object or the
source-to-sink obligation.  We count both as faithful because both preserve
the defining object and action roles.

Table~\ref{tab:alignment}(b) summarizes the primary causes of the 13
mismatches.  Action selection accounts for six: the rule chooses the wrong
source, discharge, or sink operation.  Four others track the wrong object or
lifetime, often by selecting a neighboring owner or field.  The remaining
three change a defining failure predicate or the transition topology.  Object
and action selection therefore account for 10 of 13 mismatches, matching the
generation challenges made explicit by \(O\) and \(E\) rather than failures in
the shared alias or path implementation.

We further examine how semantic alignment relates to report quality.  We group
the reviewed, deduplicated findings according to the rules that reported them.
We place a finding in the aligned group if every rule that reported it is
aligned; otherwise, we place it in the misaligned group.  One finding was
reported by both aligned and misaligned rules, and we count it as misaligned.
The 13 misaligned rules produced 340 deduplicated findings but only one TP,
while aligned rules accounted for 1,131 of the 1,132 model-specific TPs.  Thus,
nearly all TPs came exclusively from seed-aligned rules.

\begin{rqanswer}
\textbf{Answer to RQ3.}
Across the three models, 236 of 249 generated rules (94.8\%) preserve the core
tracked-object relation and defining source, discharge, and sink roles of their
seed fixes.  The 13 mismatches are localized to object identity, failure
predicates, action bindings, and state topology.  Aligned rules account for
1,131 of 1,132 model-specific TPs, indicating a strong association between
seed fidelity and true-positive findings.
\end{rqanswer}

\section{Discussion}
\paragraph{A growing rule ecosystem.}
\tool turns historical fixes into persistent, executable rules that can evolve with the kernel. Rules produced from different patches or models share the same interface and backend, so a new wrapper model, alias transfer, or path
predicate can benefit the entire rule pool. Developer-confirmed findings and their fixes can become new seeds, allowing each analysis cycle to expand the available defect knowledge. Future work can further improve seed quality by
combining patch series and follow-up fixes, and extend the rule representation to cross-interface protocols, numerical constraints, and concurrent histories.

\begin{table}[t]
\caption{Seed fidelity and misalignment causes.}
\label{tab:alignment}
\centering
\footnotesize
\begin{subtable}{\columnwidth}
\centering
\caption{Seed fidelity.}
\label{tab:alignment-fidelity}
\begin{tabular*}{\columnwidth}{@{\extracolsep{\fill}}lrrrr@{}}
\toprule
Model & Rules & Aligned & Misaligned & Fidelity \\
\midrule
GPT-5.5 & 91 & 88 & 3 & 96.7\% \\
DeepSeek-v4-pro & 74 & 70 & 4 & 94.6\% \\
Opus 4.8 & 84 & 78 & 6 & 92.9\% \\
\bottomrule
\end{tabular*}
\end{subtable}

\medskip
\begin{subtable}{\columnwidth}
\centering
\caption{Primary misalignment causes.}
\label{tab:alignment-causes}
\begin{tabularx}{\columnwidth}{@{}lXr@{}}
\toprule
Cause & Differing rule decision & Rules \\
\midrule
Object identity & tracked object or lifetime & 4 \\
Failure predicate & guard domain or branch polarity & 1 \\
Action binding & source, discharge, or sink operation & 6 \\
State topology & start, reset, or violation transition & 2 \\
\bottomrule
\end{tabularx}
\end{subtable}
\end{table}

\paragraph{Orthogonal report verification.}
RQ2 evaluates initial report pools before report-level postprocessing.
LLM-based contextual analysis and triage can filter static-analysis alerts and
refine overly broad checkers~\cite{iris,knighter}, while candidate-level
validation supports rule-guided bug discovery~\cite{bugstone}.  A downstream
agent could inspect each report's seed, bound actions, object history, and
source context to reject infeasible or unrelated candidates without changing
generation or execution.

\paragraph{Artifact Availability.}
To support artifact evaluation and reproducibility, we provide a project
repository containing the rule synthesizer, shared backend, 100 historical
repair commits, three frozen rule pools with their scope maps, and report
deduplication code.  The accompanying Docker image includes the Linux v6.16
source tree, LLVM IR, and compilation database needed to reproduce the scans.
The artifact is available at
\url{https://github.com/THU-Agent/TyPatch}.

\section{Related Work}
\label{sec:related}

\subsection{Traditional Static Analysis}

The Linux kernel has long used static analysis to assist code review and
defect detection.  Coccinelle uses the Semantic Patch Language to describe
patterns over syntax and control flow; Smatch provides an extensible rule
interface for C and the Linux kernel; and the Clang Static Analyzer uses
path-sensitive, interprocedural symbolic execution to let checkers observe
program callbacks, maintain abstract states, and generate
reports~\cite{padioleau2008coccinelle,lawall2018coccinelle,smatch,clangsa}.
These tools provide
infrastructure for applying existing rules at scale, but new defect knowledge
must still be encoded as corresponding rules.

Earlier general and kernel-specific analyzers established system-rule checking,
path-sensitive verification, and scalable bug finding~\cite{engler2000metal,
mops,ball2001slam,das2002esp,xie2005saturn,calcagno2015infer,chou2001errors,
drchecker,kminer,unisan,dcuaf,fitx}.

Object-lifecycle constraints commonly require typestate, alias, and path
analysis.  Typestate associates an abstract state with a program object and
updates that state in response to relevant actions; an operation that is not
permitted in the current state constitutes a
violation~\cite{strom1986typestate,deline2001vault,fink2006typestate,
bierhoff2007typestate,crysl,cognicrypt}.
PATA develops path-sensitive and alias-aware typestate analysis for
operating-system code, while SPATA improves the scalability of object and
state propagation through interprocedural alias summaries~\cite{pata,spata}.
These systems show that accurately executing state rules requires specialized
handling of object identity, path feasibility, and interprocedural propagation.

Another line of work automatically recovers rules from common behaviors,
execution traces, revision histories, and source-level API
usages~\cite{ammons2002mining,dynamine,perracotta,jadet,grouminer,deepmining,
engler2001deviant,prminer,apisan}.  Spinfer instead infers semantic patches from
multiple code-change examples~\cite{spinfer}.  These approaches reduce manual
rule construction but typically require many similar examples or restrict
rules to predefined code patterns.

Historical bug fixes provide a more direct source of knowledge.  Known
vulnerable code and human fixes have been used to find unpatched clones and
learn repair patterns~\cite{redebug,vuddy,getafix}.  APHP jointly
analyzes patch code and commit descriptions to extract API post-handling
specifications consisting of a target API, critical variable, post-operation,
and path condition, and then applies corresponding path analyses to detect
missing checks and resource-handling bugs~\cite{aphp}.  APHP shows that commit
descriptions can provide repair intent beyond the code change itself, although
its specifications primarily describe checks or paired operations that should
follow an API call.

SEAL compares the program-dependence graphs before and after a security fix
and infers Linux interface specifications from changed interprocedural
value-flow paths.  Its specifications center on source-to-use reachability,
path conditions, and use-site order, and are executed by a shared value-flow
analysis~\cite{seal}.  \tool differs from SEAL in both the evidence used for
rule construction and the resulting representation.  SEAL derives value-flow
relations from program-dependence changes.  \tool jointly uses the commit
message, code diff, and source context to generate an explicit description
of the tracked object, action, guard, state, and transition of a typestate rule.
State constraints expressed in a commit description or API contract, but not
necessarily visible as a changed value-flow path, can therefore participate in
rule construction.

We attempted to include SEAL in the quantitative evaluation, but encountered
practical difficulties: its publicly available version depends on private
commercial tools and is incompatible with the recent Linux kernel version used
in our experiments.  \knighter reports the same
limitations~\cite{seal,knighter}.

\subsection{LLM-Assisted Static Analysis}

One line of work uses LLMs to provide semantic information to an existing
static analyzer.  LLift integrates LLM reasoning into a conventional analysis
workflow for practical bug detection.  IRIS uses an LLM to infer taint sources
and sinks and integrates these specifications into whole-repository CodeQL
analysis.  LAMeD generates annotations for allocation and deallocation
functions, which are then consumed by an existing memory-leak
analyzer~\cite{llift2024,iris,lamed}.  These
approaches use LLM-supplied API models, annotations, or other semantic
information within a predefined analysis task.

Other systems use patch-derived rules to locate candidates and then invoke an
LLM to judge each candidate.  SpecAuditor extracts audit specifications from
historical patches to locate relevant code and guide model-based auditing.
BugStone summarizes a recurring error pattern from a repaired instance, uses
program analysis to retrieve structurally similar candidates, and then asks
an LLM whether each candidate shares the same root
cause~\cite{specauditor,bugstone}.  These systems use explicit rules to reduce
the LLM's search space, but the scan still requires LLM reasoning over
individual candidates.

\knighter is the work most related to \tool.  It summarizes a bug
pattern from a Linux repair patch, generates a checker implementation plan,
and uses an LLM to implement a complete checker from CSA templates and helper
functions.  The generated code undergoes compilation repair and is
evaluated on the buggy and fixed versions of the patch~\cite{knighter}.
\knighter demonstrates kernel-scale bug finding with LLM-generated,
path-sensitive checkers.

Complete-checker generation offers strong expressiveness because callbacks,
abstract states, object propagation, condition handling, and report logic can
all be customized for an individual patch.  At the same time, the model output
contains both patch-specific defect semantics and general analyzer
implementation.  Templates, few-shot examples, and helper functions reduce
framework-level burden, but each generated checker still composes its own
callbacks, object representations, path states, and report logic.

\tool changes what the model must produce.  Instead of implementing a complete
checker, the model describes which object to track and which program events
change its state or constitute a violation.  The shared backend binds this rule
to the program and carries the object's state through aliases and control flow.
All rules share analysis semantics, and generation errors remain confined
to the rules.

\section{Conclusion}

This paper introduces \tool, which generates structured typestate rules from
Linux kernel patches and executes them with a shared alias-aware,
path-sensitive backend.  Reusing rather than regenerating the analyzer makes
construction more reliable and substantially cheaper.
Across 100 fixes and three models, \tool finds 559 distinct bugs in Linux
v6.16, including 121 confirmed by kernel developers, while 236 of 249 rules
preserve their seed's core defect relation.

On the matched 38 patches, \tool uses 88.3--90.1\% fewer generation
tokens than \knighter while producing more artifacts and more precise initial
reports under both models.  Structured rule generation therefore offers
stable, low-cost, scalable analysis that grows with repair history.

\section*{Acknowledgments}
Generative AI systems were used as experimental subjects in the reported
rule-generation workflows and to assist language editing.  The authors verified the
paper's technical claims, code, and experimental results.

\bibliographystyle{ACM-Reference-Format}
\bibliography{references}

\appendix

\section{Supplementary Details}
\label{app:details}

\subsection{Patch-to-Rule Algorithm}
\label{app:construction}

Algorithm~\ref{alg:synthesis} summarizes evidence construction, structured
generation, the two-stage validation pipeline, and field-level repair.

\begingroup
\renewcommand{\thealgocf}{A.1}
\begin{algorithm}[!ht]
\caption{Patch-to-Rule Construction}
\label{alg:synthesis}
\DontPrintSemicolon
\KwIn{Repair commit \(p\), kernel source tree \(\mathcal{K}\), rule schema \(\mathcal{S}\)}
\KwOut{Accepted typestate rule \(R_v\), or \textsc{NoRule}}
\BlankLine
\AlgComment{Evidence construction}\;
\(c \leftarrow \operatorname{BuildContext}(p.\mathit{message},p.\mathit{diff},\mathcal{K})\)\;
\AlgComment{Structured rule generation}\;
\(y \leftarrow \operatorname{GenerateRule}(c,\mathcal{S})\)\;
\AlgComment{Two-stage validation and repair}\;
\For{\(i \leftarrow 0\) \KwTo \(2\)}{
  \If{\(y=\mathrm{NoRule}\)}{\KwRet \textsc{NoRule}\;}
  \(F_g \leftarrow \emptyset\); \(F_t \leftarrow \emptyset\); \(F_b \leftarrow \emptyset\)\;
  \((R_c,F_s) \leftarrow \operatorname{ValidateSchema}(y,\mathcal{S})\)\;
  \If{\(F_s=\emptyset\)}{
    \(F_g \leftarrow \operatorname{GroundBindings}(R_c,c,\mathcal{K})\)\;
  }
  \If{\(F_s\cup F_g=\emptyset\)}{
    \(R_v \leftarrow \operatorname{NormalizeRule}(R_c,p.\mathit{diff})\)\;
    \(F_t \leftarrow \operatorname{ValidateStateMachine}(R_v)\)\;
  }
  \If{\(F_s\cup F_g\cup F_t=\emptyset\)}{
    \(F_b \leftarrow \operatorname{ValidateBackend}(R_v)\)\;
    \If{\(F_b=\emptyset\)}{
      \(\operatorname{SerializeTS}(R_v)\)\; \KwRet \(R_v\)\;
    }
  }
  \If{\(i=2\)}{\KwRet \textsc{NoRule}\;}
  \(y \leftarrow \operatorname{RepairRule}(c,y,F_s\cup F_g\cup F_t\cup F_b)\)\;
}
\KwRet \textsc{NoRule}\;
\end{algorithm}
\endgroup

\subsection{Implementation Details}

Patch-to-rule construction is implemented in Python, and the shared backend is
implemented in C++ over LLVM IR.  The construction context includes
at most three changed-function bodies,
up to 200 lines each and 500 lines in total.  Source expansion resolves
ordinary functions, multi-line macros, one-level wrappers, and kernel
\code{DEFINE_FREE} cleanup declarations, and follows one level of callees from
those definitions.  Every excerpt retains \path{file:line} provenance and is
available to the grounding checks after generation.  For few-shot prompting,
we select at most two patch-to-rule examples by family features from a frozen
set whose rules have already executed on the backend; no example is supplied
when no family match exists.

The model-facing schema contains \(R\).  Its state domain records declared join
cases, while the \(\Phi\) part of \(K\) records report constraints.
A deterministic serializer completes \(J_Q\), maps \(\Phi\) to the backend's
fail-open path-screening policy, emits the \code{.ts} format, and attaches the selected
scan settings.  When a patch names an allocation or release function, context
construction may also provide related kernel functions for the model to
consider.  The generated IR records the functions selected by the model, and
the normalized \code{.ts} rule records the executable bindings.

\subsection{Evaluation Protocol}

\paragraph{Rule scope and scan configuration.}
\label{app:scope}
Some patch-derived rules depend on module-specific API or lifecycle
conventions.  We therefore assign each rule either kernel scope or module
scope before scanning.  Kernel scope covers all link targets in the
compilation database, whereas module scope selects a subset of those targets.
The initial scope is derived from the source distribution of function
bindings in the generated and executable rules.  Generic or unresolved
bindings lead to kernel scope.  An LLM then retains or widens an initially
local scope, but cannot narrow it.  Rules with identical target sets are
executed together to reduce unnecessary rule--target combinations.
Both RQ1 and RQ2 enable this step for \tool.

\paragraph{Report collection and review.}
We used a two-hour limit per translation unit.  RQ1 merges reports by source
site and its connected downstream sinks.
Multiple sinks reached from one source form one finding; components from
different rules or object keys are joined transitively when they share a
sink.  Reports at the original seed sites and in kernel test code are excluded
from production-bug totals.  To keep exhaustive source review feasible, the
RQ1 report pools exclude reports from any rule whose whole-kernel scan produces
more than 2,000 raw reports.  This removes 2 of 91 GPT rules, 1 of 74 DeepSeek
rules, and 3 of 84 Opus rules.  RQ2 applies no report-count gate.

The matched set contains the 38 commits in \knighter's public collection that
belong to the six sequential object-state families represented by both
systems: 6 null-dereference, 5 leak, 7 use-after-release, 8 double-free, 5
uninitialized-data, and 7 misuse fixes.  RQ2 deduplicates separately within
each rule or checker by sink file, function, and line; reports from different
generated artifacts remain distinct.  We use \knighter commit
\code{f4e834b30741} and its published default configuration.  \tool follows
the construction procedure in Section~\ref{sec:construction}; all generation
tokens from both workflows are counted.  Report refinement and triage are
outside both artifact-generation arms.

All reported populations were manually reviewed against the same Linux v6.16
source.  A report counts as a true positive only when its ordered actions
operate on the same abstract object along a feasible path.

\balance
\paragraph{Model-usage accounting.}
For \tool, the generation-token totals reported in RQ1 and RQ2 include
initial rule generation, all repair attempts, and LLM-based scope planning;
the reported model-call counts likewise include scope-planning calls.
Token totals use provider-reported input, cached-input, and output counts.
Backend execution uses the saved scope maps without further model calls,
and manual review does not consume model tokens.

\subsection{Additional Results}
\label{app:additional-results}

\begin{table}[H]
\caption{RQ1 rule-construction cost, in thousands of model tokens.  Totals
include initial generation, repair attempts, and scope planning.}
\label{tab:rq1tokens}
\centering
\footnotesize
\begin{tabular}{@{}lrrrr@{}}
\toprule
Model & Rules & Input & Output & Total \\
\midrule
GPT-5.5 & 91 & 2,229 & 294 & 2,523 \\
DeepSeek-v4-pro & 74 & 1,955 & 956 & 2,911 \\
Opus 4.8 & 84 & 2,964 & 152 & 3,116 \\
\bottomrule
\end{tabular}
\end{table}

\begin{table}[H]
\caption{RQ2 true-positive report instances by family.}
\label{tab:matched38families}
\centering
\footnotesize
\begin{tabularx}{\columnwidth}{@{}>{\raggedright\arraybackslash}Xrrrr@{}}
\toprule
& \multicolumn{2}{c}{GPT-5.5} & \multicolumn{2}{c}{DeepSeek-v4-pro} \\
\cmidrule(lr){2-3}\cmidrule(l){4-5}
Family & \tool & \knighter & \tool & \knighter \\
\midrule
Memory/Resource Leak & 138 & 1 & 141 & 7 \\
Use After Release & 30 & 2 & 0 & 6 \\
Uninitialized Data & 32 & 4 & 0 & 20 \\
Null-Pointer Dereference & 14 & 7 & 17 & 12 \\
Double Free & 0 & 1 & 0 & 3 \\
Misuse & 0 & 0 & 0 & 10 \\
\midrule
\textbf{Total} & \textbf{214} & \textbf{15} & \textbf{158} & \textbf{58} \\
\bottomrule
\end{tabularx}
\end{table}

Table~\ref{tab:matched38families} gives the family-level composition behind
the aggregate RQ2 results.  Because reports from different artifacts remain
distinct, these entries are report instances rather than distinct bugs.

Within \tool, DeepSeek-v4-pro achieves higher precision than GPT-5.5
(9.11\% vs.\ 4.67\%).  The difference is dominated by three overly broad
GPT-5.5 rules, which produce 2,729 of the 2,849 additional reports relative to
DeepSeek-v4-pro.  One rule
approximates a concurrent use-after-release as a sequential free--use relation
and produces 1,753 reports; two uninitialized-data rules produce 545 and 431.
Their false positives arise primarily from different-object propagation,
infeasible paths, and guards or initialization that precede the reported use.

\end{document}